\documentclass[sigconf]{acmart}
\pdfoutput=1

\usepackage{amssymb}
\usepackage{array}
\usepackage{url}
\usepackage{dblfloatfix}
\usepackage{booktabs}
\usepackage{amsmath}
\usepackage{tabularx}
\usepackage{colortbl}
\usepackage{hyperref}

\usepackage{graphics}
\usepackage{epsfig}
\usepackage{multirow}
\usepackage{flushend}

\usepackage{url}
\usepackage{tikz}
\usepackage{enumitem}
\usepackage{multicol}

\usepackage{verbatim}
\usepackage{float}
\usepackage{subcaption}

\usepackage{algorithm}
\usepackage{algpseudocode}

\usepackage{marginnote}
\usepackage{xspace}

\newcommand{\HARD}{\texttt{DL-HARD}\xspace}

\newcommand{\squishlist}{
 \begin{list}{$\bullet$}
  { \setlength{\itemsep}{0pt}
     \setlength{\parsep}{1pt}
     \setlength{\topsep}{1pt}
     \setlength{\partopsep}{0pt}
     \setlength{\leftmargin}{1.5em}
     \setlength{\labelwidth}{1em}
     \setlength{\labelsep}{0.5em} } }

\newcommand{\squishend}{
  \end{list}  }

\copyrightyear{2021} 
\acmYear{2021} 
\setcopyright{acmlicensed}\acmConference[SIGIR '21]{Proceedings of the 44th International ACM SIGIR Conference on Research and Development in Information Retrieval}{July 11--15, 2021}{Virtual Event, Canada}
\acmBooktitle{Proceedings of the 44th International ACM SIGIR Conference on Research and Development in Information Retrieval (SIGIR '21), July 11--15, 2021, Virtual Event, Canada}
\acmPrice{15.00}
\acmDOI{10.1145/3404835.3463262}
\acmISBN{978-1-4503-8037-9/21/07}

 \author{Iain Mackie}
 \affiliation{
       \institution{University of Glasgow}
        \city{Glasgow, Scotland, UK}
 }
 \email{i.mackie.1@research.gla.ac.uk}

 \author{Jeffrey Dalton}
 \affiliation{
       \institution{University of Glasgow}
        \city{Glasgow, Scotland, UK}
 }
 \email{jeff.dalton@glasgow.ac.uk}

 \author{Andrew Yates}
 \affiliation{
       \institution{Max Planck Institute for Informatics}
        \city{{Saarbr\"u}cken, Germany}
 }
 \email{ayates@mpi-inf.mpg.de}

\renewcommand\footnotetextcopyrightpermission[1]{} 
\begin{document}
\fancyhead{}

\title{How Deep is your Learning: the \HARD\ Annotated Deep Learning Dataset}

\begin{abstract}

Deep Learning Hard (\HARD) is a new annotated dataset designed to more effectively evaluate neural ranking models on complex topics. It builds on TREC Deep Learning (DL) topics by extensively annotating them with question intent categories, answer types, wikified entities, topic categories, and result type metadata from a commercial web search engine. Based on this data, we introduce a framework for identifying challenging queries. \HARD contains fifty topics from the official DL 2019/2020 evaluation benchmark, half of which are newly and independently assessed. We perform experiments using the official submitted runs to DL on \HARD and find substantial differences in metrics and the ranking of participating systems. Overall, \HARD is a new resource that promotes research on neural ranking methods by focusing on challenging and complex topics.

\end{abstract}

\maketitle
\section{Introduction}
\label{sec:intro}

The development of new machine learning models for ranking is an important area of Information Retrieval research, with a recent emphasis on neural language models \cite{yates2021pretrained}. These language models are state-of-the-art for both retrieval \cite{nogueira2019passage, nogueira2020document, li2020parade} and natural language understanding tasks \cite{devlin-etal-2019-bert, raffel2020exploring, NEURIPS2020_1457c0d6}. They are used by leading commercial web search engines to improve ranking and question answering (QA) effectiveness.\footnote{\url{https://blog.google/products/search/search-language-understanding-bert/}} The focus of this resource is to support the measurement of progress on challenging ranking topics where these new classes of models fail.

The MS MARCO leaderboard is a leading benchmark for both passage and document ranking.  It uses real web queries that are candidates from Bing's web QA system. Neural language models have made significant improvements on these types of question-intent queries due to their longer natural language nature. 

\begin{figure}[h]
\centering
\includegraphics[scale=0.189]{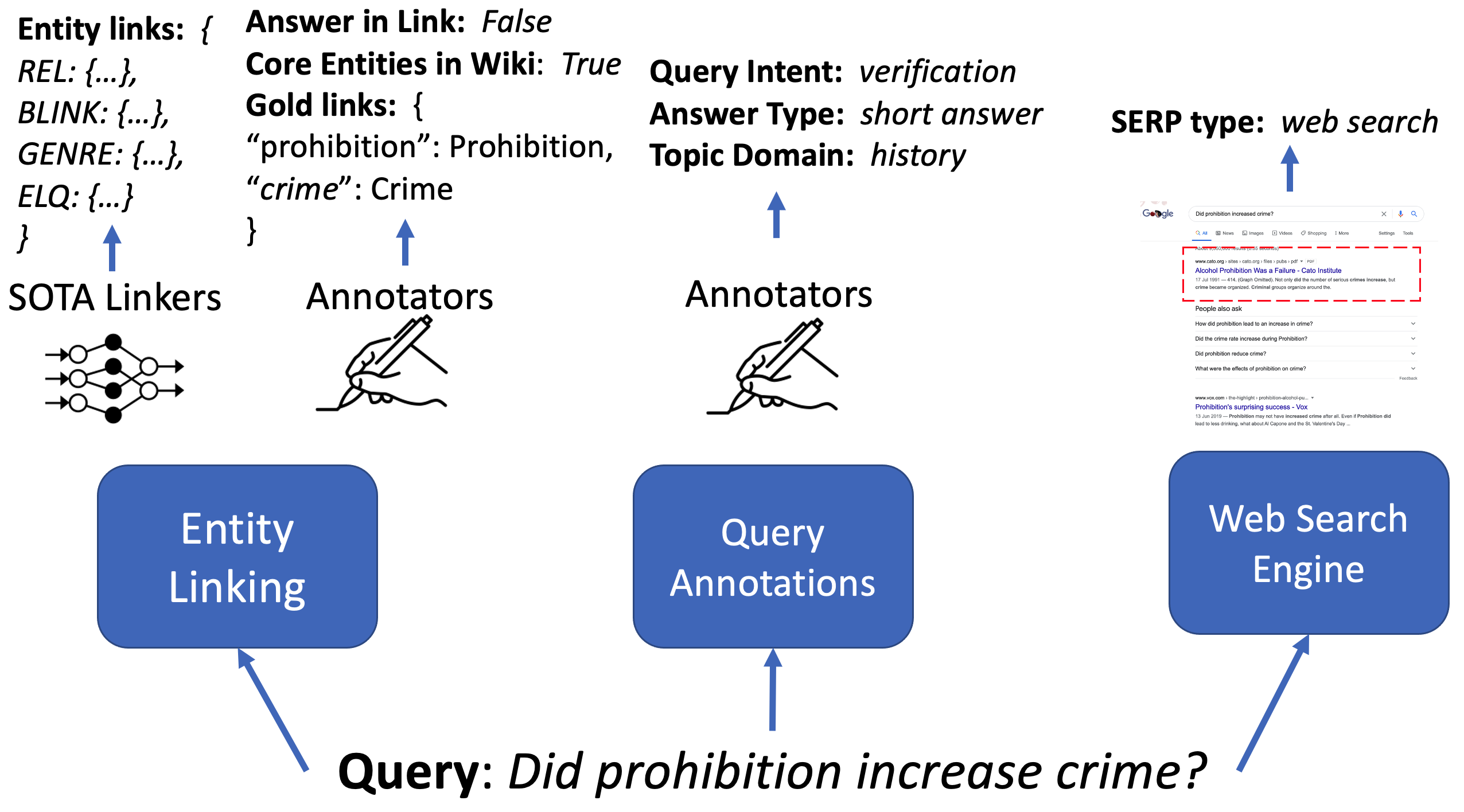}
\caption{\HARD annotation process overview.}
\label{fig:dl_hard_annotations}
\end{figure}

MS MARCO contains many queries with sparse relevance labels, whereas the TREC Deep Learning track provides a smaller subset assessed more deeply by professional assessors. The large volume of sparse MS Marco data for training, and the high-quality NIST judgments for evaluation, result in DL being a significant step forward for the community. As part of the publicly available \HARD dataset, we augment DL with rich manual and automatic query annotations on all four hundred queries (assessed and not assessed). These rich annotations include Question Intent Types \cite{Cambazoglu2021AnIT}, our own specially developed answer types, result types from a leading web search engine, coarse topic categories, and automatic and gold entity mentions linked to Wikipedia (Figure \ref{fig:dl_hard_annotations}). Such annotations enable developing new methods for identifying `hard' queries to inform future benchmark construction.

Using DL to test modern entity-centric and neural ranking algorithms for long documents is challenging. First, the reported system effectiveness is relatively high, even for existing baseline systems, with a median mean reciprocal ranking above 0.8 for both the DL 2020 document ranking and passage ranking tasks. This appears to show that current deep learning methods leave little headroom for improvement. This work demonstrates that this is not the case in practice, motivating the need for a resource that builds on proven DL data and provides the headroom required for modern systems. Second, the DL queries vary in terms of difficulty, intent, answer type, etc. What are the `right' queries to focus on when evaluating state-of-the-art neural models? Current commercial web search engines are already tuned to rapidly answer many diverse queries. Based upon studying web search engine behaviour on DL, we find that a significant proportion of the queries are `solved'. For example, many are factoid questions that can be answered from Knowledge Graphs or with lookups from structured data sources.  

\begin{figure}[h]
\centering
\includegraphics[scale=0.172]{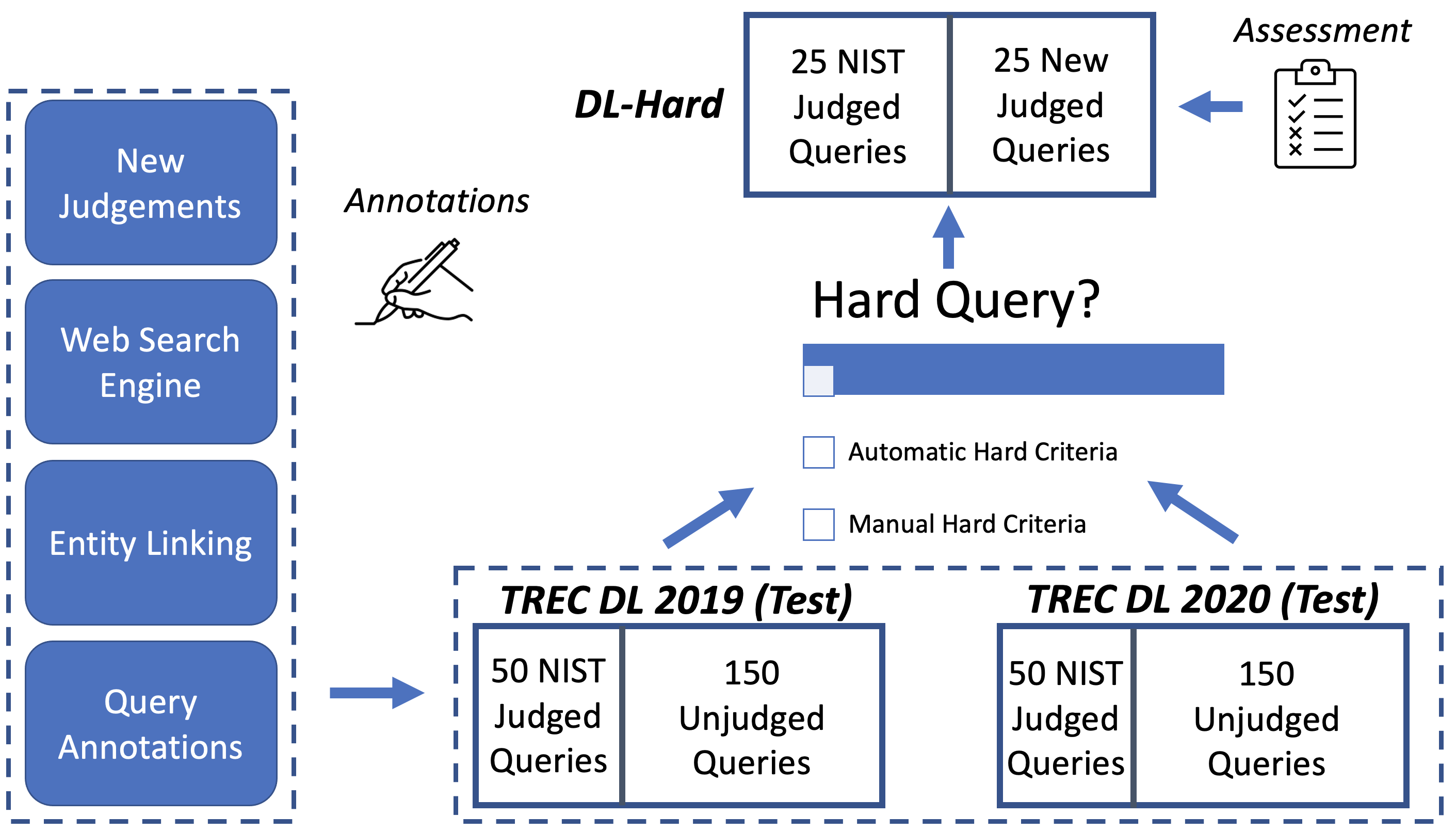}
\caption{\HARD dataset overview.}
\label{fig:dl_hard_overview}
\end{figure}

To address these issues, we introduce the \HARD dataset\footnote{\HARD is available at \url{https://github.com/grill-lab/DL-HARD}} that consists of `hard' DL topics from 2019 and 2020. \HARD provides annotations on the full four hundred queries and a benchmark consisting of the fifty most difficult queries across both years. These include twenty-five previously assessed queries and twenty-five queries with new sparse judgments annotated at a passage and document level (Figure \ref{fig:dl_hard_overview}).

We perform an empirical evaluation of \HARD topics on all the submitted runs to DL 2020. We find that \HARD topics are substantially more difficult and lead to swaps in the ranking of systems. When considering the twenty-five queries assessed by DL, each system moved on average 4.6 places in the overall system ranking. For the twenty-five queries with new judgments, this results in even larger changes to the relative ordering of systems. We note that \HARD queries are more complex and contain a higher fraction of list and long answer results than DL.

The key contributions of this resource include:
\begin{itemize}

\item Diverse manual and automatic annotations for all DL topics: intent type, answer type, search engine result type, topic category, and wikified entities.

\item A new \HARD benchmark with specified test folds for both document and passage tasks. Half the queries are newly assessed.

\item A semi-automatic method for identifying challenging queries that leverages evidence from commercial web search systems, query intent, and other metadata. 
 
\item A study of the behavior of official DL submissions on the new \HARD queries.

\end{itemize}

\section{Related Work}
\label{sec:related}

The MS MARCO passage and document collections \cite{nguyen2016ms} consist of queries, web passages or documents, and sparse relevance judgments between them. This dataset derives passage-level relevance judgments from the MS MARCO Question Answering dataset by treating any passage containing a correct answer for the query to be relevant. These passage-level judgments are transferred to the document level by labelling any document containing a relevant passage to also be relevant. While this label transfer approach requires little manual effort and enables the creation of a large dataset, it has several issues that may make it artificially easy. First, all queries have an associated answer from the QA dataset. Second, all queries have a single passage answer. Finally, the set of documents is limited in scope to ones that are a passage candidate for one of the queries.

The queries in MS MARCO are longer than typical web queries (5.8 words on average across DL), which are more challenging to handle than short keyword queries~\cite{huston2010evaluating}. However, many are factoid or begin with `wh-' words, which may be easier than more open-ended long queries.

The TREC Deep Learning (DL) track \cite{Craswell:2003.07820:2020,craswell2021overview} ran tasks using the MS MARCO passage and document collections. Assessments by NIST annotators address the label sparsity issue by providing judgments pooled to a greater depth. We address the query difficulty issue by building upon DL and identifying the most complex queries. 
In this work, we define web query answer types with a new taxonomy developed bottom-up for MS MARCO to help categorise challenging and interesting topics. The developed `hard criteria' helps to select these topics systematically.  

Cambazoglu et al. \cite{Cambazoglu2021AnIT} study the types of queries in MS MARCO. They create a taxonomy of intents for questions, the types of named entities present in questions, the types of question words used, and the answer's expected granularity. We manually annotate all DL queries with intents from their taxonomy and additionally introduce a complementary schema that is more fine-grained. Similarly, the task of determining a question's answer type has been widely studied in QA. For example, by associating salient terms in the question, such as `wh-' words, with answer types identified in a corpus \cite{10.5555/1289189.1289276}.

Our dataset also complements previous work in QA, which generally refers to the task of identifying a relevant text span in response to a question. The TREC Question Answering track \cite{Voorhees_TREC2001} constructed a series of QA benchmarks beginning in the late 1990s. 
More recently, the SQuAD \cite{rajpurkar-etal-2016-squad} and Natural Questions \cite{47761} benchmarks each provide over 100,000 crowdsourced questions and associated answer spans. In contrast to seeking factoid or short answers, \HARD dataset focuses on complex answers that can be long and multi-faceted.

\section{Task and Judgments}
\label{sec:task-judgments}

We now describe the resource task and relevance assessment for the \HARD\ dataset. 

\textbf{Task Definition.} The task is an information-seeking passage and document ranking task that follows the one described in the TREC Deep Learning track \cite{Craswell:2003.07820:2020}. Because the use case for \HARD\ emphasizes challenging ad-hoc retrieval, the query intents are more likely long descriptions, multiple answers, a list, or require reasoning. The criteria used to filter queries include: spelling errors, incomplete, ambiguous, or target a specialized structured vertical, i.e. calculator, maps, weather, or dictionary. Additionally, since factoid QA is already a well-studied area in TREC \cite{Voorhees_TREC2001} and the NLP community \cite{rajpurkar-etal-2016-squad}, \HARD\ queries are primarily non-factoid.

For the experimental setup, we provide five pre-defined `standard' folds to be used for k-fold cross-validation. The dataset could also be used purely as a test set in a zero-shot capacity. \HARD\ can evaluate end-to-end retrieval or re-ranking leveraging provided baseline runs.     

\textbf{Relevance Assessment Process.} The resource uses the full NIST assessments for previously judged topics. There are also new passage and document level judgments provided for unjudged queries from DL. 

We perform relevance assessment on a graded scale using the same guidelines as the original track for the new judgments. We assess passages returned in the MS MARCO QA corpus and the documents they are drawn from. Unlike the MS MARCO sparse judgments, which generally include only one relevant passage per query, we assess all of the top ten responses. Experienced IR researchers (the authors) perform the annotations. 

To calculate agreement with the NIST assessors, we additionally judge the top QA passage responses for 24 queries from DL (12 from each year). We find Krippendorff's alpha is 0.47 on the passage judgments for these queries and 0.43 on the document judgments, which indicates moderate agreement. Krippendorff's alpha drops to 0.12 when transferring passage assessments to documents, illustrating the difficulty of automatically transferring passage assessments. For this reason, we adopt document-level relevance judgments for the official \HARD document ranking task.

\section{Annotations}
\label{sec:annotations}
We detail the annotations provided within the resource. We use these annotations in Section \ref{sec:hard-criteria} to develop the criteria for selecting hard topics for the \HARD dataset.

\subsection{Question Intent Annotation}
We apply the question intent taxonomy developed for MS MARCO web questions \cite{Cambazoglu2021AnIT}. In contrast to other taxonomies, this has a more fine-grained taxonomy developed bottom-up for MS MARCO. We use their \textit{Query Intent Categories} and guidelines to annotate all official DL queries. At least one author performs each annotation, and ambiguous instances resolved by majority vote. To our knowledge, this is the first resource to make these annotations publicly available. 

The distribution of the query intents on the complete DL queryset as well as \HARD\ is shown in Table \ref{tab:query_intent:categories}. The most notable difference is the increase in List intents from DL (10.2\% across 2019 and 2020) to \HARD\ (34.7\%). The annotators note that list queries are harder as the user seeks multiple entities or facts that could span many documents. The proportion of Quantity intents in \HARD\ is much lower as most of these queries are either simple factoid-QA questions (`hydrogen is a liquid below what temperature') or highly underspecified and should be clarified (i.e. `cost of interior concrete flooring'). \HARD\ also filters out Language and Weather intents.

\begin{table}[h]
\caption{Query Intent Categories for DL and \HARD.}
\label{tab:query_intent:categories}

\begin{tabular}{|l|r|r|r|r|}
\hline
\textbf{Intent Category} & \multicolumn{1}{l|}{\textbf{DL-2019}} & \multicolumn{1}{l|}{\textbf{DL-2020}} & \multicolumn{1}{l|}{\textbf{\HARD}} \\ \hline
Attribute                & 1                                     & 5                                     & 1                                                  \\ \hline
Description              & 21                                    & 20                                    & 20                                                 \\ \hline
Entity                   & 3                                     & 4                                     & 3                                                  \\ \hline
Language                 & 0                                     & 2                                     & 0                                                  \\ \hline
List                     & 7                                     & 2                                     & 17                                                 \\ \hline
Process                  & 1                                     & 1                                     & 0                                                  \\ \hline
Quantity                 & 5                                     & 6                                     & 3                                                  \\ \hline
Reason                   & 3                                     & 4                                     & 4                                                  \\ \hline
Verification             & 1                                     & 1                                     & 2                                                  \\ \hline
Weather                  & 1                                     & 0                                     & 0                                                  \\ \hline
\end{tabular}
\end{table}

\subsection{SERP Result Types}
\label{sec:searchengine}
To retrieve the Search Engine Results Page (SERP), we manually issue every query on a Desktop browser to an English language Google search engine from the United Kingdom in `incognito mode'. The authors inspect the results and save the raw HTML content to include as part of the resource. We note queries with potential localization issues (local store phone number, location, or hours). Based on the criteria described in Section \ref{sec:hard-criteria}, we exclude these queries because they are unanswerable without local context (not provided in DL).

For each query, we annotate the type of rich results returned in the SERP and whether the Knowledge Graph \cite{Noy2019IndustryscaleKG} is used (the raw HTML shows the schema elements). Although many possible types of rich results may be present in a SERP, the ones highlighted below are the most prevalent for DL queries:
\begin{itemize}
  \item \textit{Spell correct or suggestion}: Shows a suggested spelling correction or alternative query.
  \item \textit{Knowledge Graph (KG)}: Returns a specific answer entity, list of entities, or their attributes from structured entity data. This includes media structured results for television, movie, and music entity information.
  \item \textit{Dictionary}: Provides a dictionary definition of one or more words.
  \item \textit{Weather}: Shows the weather forecast for a locale via an embedded panel.
  \item \textit{Map}: Shows a Maps vertical result, optionally with possible driving directions.
  \item \textit{Web Short Answer}: Shows a specific string short answer, possibly with a separate supporting evidence passage from a web result. 
  \item \textit{Web Passage}: Shows a passage (or portion of a list or table) from a web result. It may highlight possible answers.
  \item \textit{Web Search}: Shows a standard list of `10 blue links'.  
\end{itemize}

The distribution of the response types for DL assessed and \HARD\ topics is shown in Table \ref{search_engine:categories}. By far, the most frequent response type is a Web Passage, which is unsurprising given that the queries are questions originally used for QA. It shows that over 20\% of the queries are answered directly with short factoid answers, with 12.5\% of results from a structured source.  Although the Google answer quality is not explicitly assessed, we observe only 2 instances of clear failure due to imprecise and/or ambiguous queries. This indicates that existing models (neural or otherwise) can adequately satisfy these `easy' factoid queries.

\begin{table}[h]
\caption{SERP result types distribution for DL Track and \HARD.}

\label{search_engine:categories}

\begin{tabular}{|l|r|r|r|}
\hline
\textbf{SERP Result} & \multicolumn{1}{l|}{\textbf{DL-2019}} & \multicolumn{1}{l|}{\textbf{DL-2020}} & \multicolumn{1}{l|}{\textbf{\HARD}} \\ \hline
Dictionary           & 1                                     & 3                                     & 1                                                  \\ \hline
KG                   & 1                                     & 5                                     & 2                                                  \\ \hline
Weather              & 1                                     & 0                                     & 0                                                  \\ \hline
Web Passage          & 24                                    & 25                                    & 28                                                 \\ \hline
Web Search           & 12                                    & 9                                     & 18                                                 \\ \hline
Web Short Answer     & 4                                     & 3                                     & 1                                                  \\ \hline
\end{tabular}
\end{table}

We evaluate whether the SERP answer type is a good heuristic for systematically identifying hard queries by mapping the SERP annotations onto 2019/2020 DL runs. The expectation is that queries within the Web Search category should be a reasonable proxy for a hard query, i.e. either (1) the search engine could not find an answer for the query, or (2) the query could not be satisfied by a short passage or entity.  Based on an analysis of assessed DL queries on the best DL systems, defined as those with above-median NDCG@10, this trend holds (Table \ref{search_engine:correlation}). Statistical analysis shows much lower NDCG@10 and Recall@100 with negative Pearson correlation coefficients. Supporting Web Search answer type as a primary feature in the `automatic hard criteria' in Section \ref{sec:hard-criteria}.

\begin{table}[h]
\caption{SERP result type performance on DL 2019/20 systems (systems above median). Mean and Pearson Correlation Coefficient (PCC) across all DL assessed queries.}
\label{search_engine:correlation}

\begin{tabular}{l|r|r|r|r|}
\cline{2-5}

                                           & \multicolumn{2}{c|}{\textbf{NDCG@10}}                                                                       & \multicolumn{2}{c|}{\textbf{Recall@100}}                                \\ \hline
\multicolumn{1}{|l|}{\textbf{SERP Result}} & \multicolumn{1}{l|}{\textbf{Mean}} & \multicolumn{1}{l|}{\textbf{PCC}} & \multicolumn{1}{l|}{\textbf{Mean}} & \multicolumn{1}{l|}{\textbf{PCC}} \\ \hline
\multicolumn{1}{|l|}{KG}                   & 0.577                              & -0.05                            & 0.794                            & 0.11                               \\ \hline
\multicolumn{1}{|l|}{Dictionary}           & 0.748                              & 0.13                            & 0.722                              & 0.04                               \\ \hline
\multicolumn{1}{|l|}{Weather}              & 0.735                              & 0.05                            & 0.464                              & -0.06                              \\ \hline
\multicolumn{1}{|l|}{Web Passage}          & 0.647                              & 0.13                            & 0.684                              & 0.04                               \\ \hline
\multicolumn{1}{|l|}{Web Search}           & 0.535                              & -0.20                           & 0.581                              & -0.16                              \\ \hline
\multicolumn{1}{|l|}{Web Short Answer}     & 0.621                              & 0.00                          & 0.731                              & 0.05                               \\ \hline
\end{tabular}
%\vspace{-6mm}
\end{table}

\subsection{Answer Type Annotations}

Previous manual and automatic annotations focus on the type of question intent or the SERP result type. Therefore, we create a new target answer type for MS MARCO web queries. The manual answer type labels are from all authors with a majority vote resolution. To develop the types, we follow a bottom-up multi-round curation similar to that used for query intents \cite{Cambazoglu2021AnIT}. The answer types are: 
\begin{itemize}
    \item \textit{Definition} - A single passage precisely and completely answers the information need. These are most commonly associated with the Description and Language query intents.
    \item \textit{Factoid} - A specific short fact answer to a question. These are often associated with Entity, Attribute, Quantity, and Location intent types.  
    \item \textit{Short answer} - A short passage (approximately a sentence) generally satisfies most information needs. Most commonly associated with Description and other factoid-like intents.
    \item \textit{Long answer} - A long passage or full document is needed to answer the query. These are associated with Description, List, and Process intents.
    \item \textit{List} - More than one answer, passage, or entity with justification is needed to answer the query.
    \item \textit{Maps} - A structured map answer is needed; this is associated with Location and local Calculation intents.
    \item \textit{Weather} - A structured weather result; corresponds to the Weather intent type.
     \item \textit{Comparison} - A comparison of two or more entities. These are associated with Description intent types.
      \item \textit{Guide} - A guide answer is a long semi-structured answer to satisfy the Process intent.
\end{itemize}

The answer types have strong associations with query intent types. However, we find that the Description intent is often quite general and does not provide guidance on the type of information needed for the answer. This is important because these answer types are useful features for topic complexity (see Section \ref{sec:hard-criteria}).

\begin{table}[t]
 \caption{Answer Type distribution for DL Track and \HARD. }
    \label{tab:topic_answer_types}
\begin{tabular}{|l|r|r|r|}
\hline
\textbf{Answer Type} & \multicolumn{1}{l|}{\textbf{DL-2019}} & \multicolumn{1}{l|}{\textbf{Dl-2020}} & \multicolumn{1}{l|}{\textbf{\HARD}} \\ \hline
Comparison             & 3                                     & 2                                     & 0                                                  \\ \hline
Definition             & 9                                     & 7                                     & 7                                                  \\ \hline
Factoid                & 12                                    & 24                                    & 5                                                  \\ \hline
Guide                  & 0                                     & 1                                     & 0                                                  \\ \hline
List                   & 9                                     & 0                                     & 15                                                 \\ \hline
Long Answer            & 6                                     & 10                                    & 13                                                 \\ \hline
Multi-Answer           & 0                                     & 1                                     & 0                                                  \\ \hline
Short Answer           & 3                                     & 0                                     & 9                                                  \\ \hline
Short Description      & 0                                     & 0                                     & 1                                                  \\ \hline
Weather                & 1                                     & 0                                     & 0                                                  \\ \hline
\end{tabular}
\end{table}

Table \ref{tab:topic_answer_types} shows the answer type breakdown for the assessed DL and \HARD\ topics. Compared with DL topics, it is clear that there are fewer Factoid responses and more List answers within \HARD.

\subsection{Query Entity Annotation}
\label{tab:entity_annotations}

Entity linking \cite{Cornolti2019SMAPHAP} and semantic parsing \cite{Berant2013SemanticPO} of question queries is an important component of modern QA systems. However, the existing DL queries do not have standard automatic or manual annotations. We provide both as part of \HARD. 

We include four state-of-the-art entity linkers developed for documents and queries: REL \cite{Hulst2020RELAE}, Blink \cite{Wu2020ZeroshotEL}, Genre \cite{decao2020autoregressive}, and ELQ \cite{Wu2020ZeroshotEL}.  We run these annotators with high-recall score thresholds, preserving score information for downstream applications, which is important for entity-based retrieval models \cite{Dalton2014EntityQF}. Based upon the automatic results, we create gold entity links to Wikipedia and metadata about the entities, i.e. (1) whether the query entity is in Wikipedia and (2) whether the Wikipedia entity satisfies the query. 

\begin{table*}[h]
\caption{Automatic hard criteria for categorising hard queries (100 labelled DL assessed topics).}\label{tab:hard_selection_criteria}
\begin{tabular}{|l|l|l|rrr}
\cline{1-3}
\multicolumn{2}{|c|}{\textbf{Include}} & \multicolumn{1}{c|}{\textbf{Exclude}} & \multicolumn{1}{l}{}                    & \multicolumn{1}{l}{}                 & \multicolumn{1}{l}{}                \\ \hline
\textbf{SERP}  & \textbf{Query Intent} & \textbf{Query Intent}                 & \multicolumn{1}{l|}{\textbf{Precision}} & \multicolumn{1}{l|}{\textbf{Recall}} & \multicolumn{1}{l|}{\textbf{F1}}    \\ \hline
Web Search     &                       &                                       & \multicolumn{1}{r|}{0.428}              & \multicolumn{1}{r|}{0.360}           & \multicolumn{1}{r|}{0.391}          \\ \hline
               & List, Reason          &                                       & \multicolumn{1}{r|}{0.588}              & \multicolumn{1}{r|}{0.400}           & \multicolumn{1}{r|}{0.476}          \\ \hline
               & List, Reason, Entity  &                                       & \multicolumn{1}{r|}{0.500}              & \multicolumn{1}{r|}{0.480}           & \multicolumn{1}{r|}{0.490}          \\ \hline
Web Search     & List, Reason          &                                       & \multicolumn{1}{r|}{0.486}              & \multicolumn{1}{r|}{0.680}           & \multicolumn{1}{r|}{0.566}          \\ \hline
Web Search     & List, Reason, Entity  &                                       & \multicolumn{1}{r|}{0.450}              & \multicolumn{1}{r|}{\textbf{0.720}}  & \multicolumn{1}{r|}{0.553}          \\ \hline
Web Search     & List, Reason, Entity  & Quantity, Weather, Language           & \multicolumn{1}{r|}{0.529}              & \multicolumn{1}{r|}{\textbf{0.720}}  & \multicolumn{1}{r|}{0.610}          \\ \hline
Web Search     & List, Reason          & Quantity, Weather, Language           & \multicolumn{1}{r|}{\textbf{0.586}}     & \multicolumn{1}{r|}{0.680}           & \multicolumn{1}{r|}{\textbf{0.630}} \\ \hline
\end{tabular}
\end{table*}

\begin{table}[t]
\caption{Topic domain category distribution for DL Track and \HARD. }
\label{tab:topic_subjects}
\begin{tabular}{|l|r|r|r|}
\hline
\textbf{Topic Domain}      & \multicolumn{1}{l|}{\textbf{DL-2019}} & \multicolumn{1}{l|}{\textbf{DL-2020}} & \multicolumn{1}{l|}{\textbf{\HARD}} \\ \hline
Business \& Finance        & 1                                     & 6                                     & 3                                                  \\ \hline
Education                  & 0                                     & 0                                     & 1                                                  \\ \hline
Entertainment \& Celebrity & 0                                     & 9                                     & 0                                                  \\ \hline
Food \& Travel             & 5                                     & 3                                     & 4                                                  \\ \hline
Health                     & 10                                    & 4                                     & 20                                                 \\ \hline
History                    & 5                                     & 6                                     & 7                                                  \\ \hline
Language \& Literature     & 1                                     & 4                                     & 2                                                  \\ \hline
Law \& Politics            & 2                                     & 2                                     & 2                                                  \\ \hline
Local                      & 1                                     & 0                                     & 1                                                  \\ \hline
Mathematics \& Science       & 13                                     & 6                                     & 10                                                  \\ \hline
Sports                     & 1                                     & 3                                     & 1                                                  \\ \hline
Technology                 & 4                                     & 2                                     & 0                                                  \\ \hline
\end{tabular}
\end{table}

\subsection{Coarse Topic Categories}

Following the categories from the TREC Conversational Assistance Track (CAsT) topics \cite{Dalton2020TRECC2}, we provide a breakdown of topics by coarse subject domain in Table \ref{tab:topic_subjects}. For 2019 we observe frequent DL topic categories to be Health, Science, and History. In 2020 there is a shift to more Entertainment (movies, tv, music), Business and Finance topics, and less Science and Health. There is also an increase in Language topics, with predominately definition queries. The largest category for \HARD is Health, a challenging category that often requires long answer responses.

\section{Hard Criteria}
\label{sec:hard-criteria}

New challenging and complex benchmark topics are required to differentiate system performance of neural ranking models. Because manually judging all candidate queries is time-consuming, we develop an `automatic hard criteria' to generate candidate queries scalably. Each candidate topic is then manually labelled using the `manual hard criteria' by multiple assessors. This process identifies 50 `hard' topics within the 400 DL topics.

\subsection{Automatic Hard Criteria}
Given that manually reviewing or creating judgments for all candidate topics is time-consuming, we explore the use of annotated metadata to generate a hard dataset. 

For simple and explainable criteria, we test explicit rule-based inclusion and exclusion filters.  We measure their agreement with the human labels and the effectiveness of existing systems on the 100 assessed DL queries (25 labelled `hard' and 75 labelled `not hard'). We present precision, recall and F1 results in Table \ref{tab:hard_selection_criteria}. We observe that the most effective rule uses Google's SERP answer type (Web Search) as a base with additional List and Reason query intents added to improve recall. We exclude intent types matching Quantity, Weather, and Language (mostly dictionary lookups). We see that adding Entity queries improves recall, but these queries also include several `easy' factoid questions. 

Although results were relatively encouraging, particularly for identifying potential hard queries, we require the `manual hard criteria' to ensure only the optimal queries are selected. Thus, the additional 25 unassessed \HARD queries (from a possible 300) combined automatic and manual criteria for labelling.  More advanced methods for automatically selecting hard queries is an area for additional future exploration.

\begin{table*}[]
\caption{Top 20 systems' effectiveness on \HARD compared with DL for the 2020 document ranking task. A complete table is available on the resource website.}
\begin{tabular}{|l|
>{\columncolor[HTML]{EFEFEF}}r |
>{\columncolor[HTML]{EFEFEF}}r |
>{\columncolor[HTML]{EFEFEF}}r |r|r|r|
>{\columncolor[HTML]{EFEFEF}}r |
>{\columncolor[HTML]{EFEFEF}}r |
>{\columncolor[HTML]{EFEFEF}}r |}
\hline
\label{tab:2020_runs_dl_vs_dl_hard_v2}

                           & \multicolumn{3}{c|}{\cellcolor[HTML]{EFEFEF}\textbf{NDCG@10}}                                                                                                                               & \multicolumn{3}{c|}{\textbf{Reciprocal Rank (RR)}}                                                                                    & \multicolumn{3}{c|}{\cellcolor[HTML]{EFEFEF}\textbf{Recall@1000}}                                                                                                                           \\ \hline
\textbf{System}            & \multicolumn{1}{l|}{\cellcolor[HTML]{EFEFEF}\textbf{\HARD}} & \multicolumn{1}{l|}{\cellcolor[HTML]{EFEFEF}\textbf{DL}} & \multicolumn{1}{l|}{\cellcolor[HTML]{EFEFEF}\textbf{\% Diff}} & \multicolumn{1}{l|}{\textbf{\HARD}} & \multicolumn{1}{l|}{\textbf{DL}} & \multicolumn{1}{l|}{\textbf{\% Diff}} & \multicolumn{1}{l|}{\cellcolor[HTML]{EFEFEF}\textbf{\HARD}} & \multicolumn{1}{l|}{\cellcolor[HTML]{EFEFEF}\textbf{DL}} & \multicolumn{1}{l|}{\cellcolor[HTML]{EFEFEF}\textbf{\% Diff}} \\ \hline
\textbf{ICIP\_run1}        & 0.452                                                      & 0.662                                                          & -21.1\%                                                       & 0.510                              & 0.736                                  & -22.7\%                               & 0.484                                                      & 0.692                                                          & -20.8\%                                                       \\ \hline
\textbf{d\_d2q\_duo}       & 0.449                                                      & 0.693                                                          & -24.5\%                                                       & 0.472                              & 0.734                                  & -26.2\%                               & 0.690                                                      & 0.842                                                          & -15.3\%                                                       \\ \hline
\textbf{fr\_doc\_roberta}  & 0.442                                                      & 0.640                                                          & -19.9\%                                                       & 0.524                              & 0.733                                  & -20.9\%                               & 0.641                                                      & 0.788                                                          & -14.6\%                                                       \\ \hline
\textbf{d\_d2q\_rm3\_duo}  & 0.438                                                      & 0.690                                                          & -25.2\%                                                       & 0.479                              & 0.735                                  & -25.6\%                               & 0.664                                                      & 0.860                                                          & -19.6\%                                                       \\ \hline
\textbf{mpii\_run2}        & 0.432                                                      & 0.613                                                          & -18.1\%                                                       & 0.468                              & 0.677                                  & -20.9\%                               & 0.484                                                      & 0.692                                                          & -20.8\%                                                       \\ \hline
\textbf{bcai\_bertb\_docv} & 0.431                                                      & 0.628                                                          & -19.6\%                                                       & 0.416                              & 0.739                                  & -32.3\%                               & 0.581                                                      & 0.760                                                          & -17.9\%                                                       \\ \hline
\textbf{ICIP\_run3}        & 0.431                                                      & 0.653                                                          & -22.2\%                                                       & 0.536                              & 0.755                                  & -21.9\%                               & 0.484                                                      & 0.692                                                          & -20.8\%                                                       \\ \hline
\textbf{bigIR-DTH-T5-F}    & 0.425                                                      & 0.591                                                          & -16.5\%                                                       & 0.559                              & 0.681                                  & -12.1\%                               & 0.581                                                      & 0.736                                                          & -15.5\%                                                       \\ \hline
\textbf{d\_rm3\_duo}       & 0.424                                                      & 0.679                                                          & -25.5\%                                                       & 0.467                              & 0.733                                  & -26.6\%                               & 0.622                                                      & 0.826                                                          & -20.4\%                                                       \\ \hline
\textbf{ndrm3-full}        & 0.415                                                      & 0.616                                                          & -20.1\%                                                       & 0.448                              & 0.716                                  & -26.8\%                               & 0.609                                                      & 0.780                                                          & -17.1\%                                                       \\ \hline
\textbf{ICIP\_run2}        & 0.413                                                      & 0.632                                                          & -21.9\%                                                       & 0.489                              & 0.733                                  & -24.4\%                               & 0.484                                                      & 0.692                                                          & -20.8\%                                                       \\ \hline
\textbf{ndrm3-re}          & 0.409                                                      & 0.616                                                          & -20.7\%                                                       & 0.455                              & 0.713                                  & -25.9\%                               & 0.484                                                      & 0.692                                                          & -20.8\%                                                       \\ \hline
\textbf{roberta-large}     & 0.408                                                      & 0.629                                                          & -22.2\%                                                       & 0.465                              & 0.739                                  & -27.4\%                               & 0.484                                                      & 0.692                                                          & -20.8\%                                                       \\ \hline
\textbf{mpii\_run1}        & 0.407                                                      & 0.602                                                          & -19.4\%                                                       & 0.494                              & 0.696                                  & -20.2\%                               & 0.484                                                      & 0.692                                                          & -20.8\%                                                       \\ \hline
\textbf{bigIR-DTH-T5-R}    & 0.407                                                      & 0.603                                                          & -19.6\%                                                       & 0.507                              & 0.697                                  & -19.0\%                               & 0.484                                                      & 0.692                                                          & -20.8\%                                                       \\ \hline
\textbf{bigIR-DH-T5-F}     & 0.404                                                      & 0.573                                                          & -16.9\%                                                       & 0.572                              & 0.659                                  & -8.8\%                                & 0.581                                                      & 0.736                                                          & -15.5\%                                                       \\ \hline
\textbf{ndrm3-orc-full}    & 0.402                                                      & 0.625                                                          & -22.3\%                                                       & 0.486                              & 0.719                                  & -23.3\%                               & 0.611                                                      & 0.784                                                          & -17.3\%                                                       \\ \hline
\textbf{ndrm3-orc-re}      & 0.397                                                      & 0.622                                                          & -22.5\%                                                       & 0.419                              & 0.692                                  & -27.3\%                               & 0.484                                                      & 0.692                                                          & -20.8\%                                                       \\ \hline
\textbf{TUW-TKL-2k}        & 0.396                                                      & 0.585                                                          & -18.9\%                                                       & 0.462                              & 0.689                                  & -22.7\%                               & 0.484                                                      & 0.692                                                          & -20.8\%                                                       \\ \hline
\textbf{TUW-TKL-4k}        & 0.393                                                      & 0.575                                                          & -18.2\%                                                       & 0.486                              & 0.694                                  & -20.8\%                               & 0.484                                                      & 0.692                                                          & -20.8\%                                                       \\ \hline
\hline
\textbf{Mean}              & \textbf{0.419}                                             & \textbf{0.626}                                                 & \textbf{-20.8\%}                                              & \textbf{0.486}                     & \textbf{0.714}                         & \textbf{-22.8\%}                      & \textbf{0.545}                                             & \textbf{0.736}                                                 & \textbf{-19.1\%}                                              \\ \hline
\end{tabular}
\end{table*}

\subsection{Manual Hard Criteria}
Hard queries are those where current models are not effective. However, not all queries where systems fail are challenging for `interesting' reasons; it could be due to missed stopwords or trivial differences in tokenization, i.e. `why did the us volunterilay enter ww1'. Additionally, under-specified queries are hard for assessors and search engines to answer definitively, i.e. `who is robert gray' (multiple Robert Grays) or `cost of interior concrete flooring' (local and ambiguous).

The authors consider both when and how systems struggle. We consider behavior in a first pass candidate retrieval (candidate recall) and second pass re-ranking (retrieval in top ranks). Queries with either type of failure are candidates for inclusion in \HARD. Each candidate topic is individually labelled and resolved across all annotators. These discussions inform the guidelines developed below:
\begin{itemize}
    \item \textit{Non-Factoid} - The query should not be answerable by a single short answer, possibly from a KG.
    \item \textit{Beyond single passage} - The query should require more than a simple definition or Wikipedia short description. 
    \item \textit{Answerable} - The topic should be answerable solely from the provided query because additional long description or narratives are not provided. Queries depending on external context should also be removed (i.e. location, temporal, etc.).
    \item \textit{Text-focused} - Queries that require non-text answers or calculation of quantities should be handled by specialized components and excluded.
    \item \textit{Mostly well-formed} - The query should not contain spelling errors or other clear language errors that would be filtered by an initial query rewriting step.
    \item \textit{Possibly Complex} - A query is desirable for inclusion if it references multiple entities, seeks a comparison, or has multiple answers.
\end{itemize}

\section{Resource Experiments}
\label{sec:exp}

We measure official TREC 2020 document run submissions on \HARD\ and compare to the original DL Track to (1) determine whether the dataset differs in system behavior and (2) measure differences in system rankings (swaps) on this dataset. For binary metrics, we consider labels of two or greater to be relevant.

The 2020 system effectiveness for DL Track, \HARD\ and the relative differences is shown in Table \ref{tab:2020_runs_dl_vs_dl_hard_v2}. On an average relative basis, \HARD\ NDCG@10 is 21.1\% lower, RR is 23.2\% lower, and Recall@1000 is 19.6\% lower. There are similar findings when evaluating the 2019 document task and shows headroom for system improvement. 

Additionally, many system swaps occur when comparing the DL Track system rankings to \HARD\ ranking. This includes a new top system (`ICIP\_run1'), and each system changes on average 4.6 places, with some systems changing as many as 12 places. This is supported by the Kendall's Tau coefficients of 0.696 (2019) and 0.641 (2020) when comparing DL Track and \HARD\ system rankings. This large number of swaps supports that removing easier queries allows for greater differentiation and more precise comparison between systems.  

Similarly, we evaluate the 2019 and 2020 DL systems on the 25 new sparse annotations using the official runs. These results cannot be directly compared to the DL Track as these queries have new judgments. Nonetheless, the top 10 systems only have an NDCG@10 of 0.314 and RR of 0.452, indicating \HARD topics with new judgments are challenging for modern systems.

\section{Conclusion}
\label{sec:con}

We introduce the \HARD\ dataset resource for evaluating modern deep learning ranking models.  It provides a challenging set of topics with new annotations: question intent types, answer types, categories, entity links, and metadata from Google SERPs. We contribute new judgments for queries not previously assessed by NIST.  All of the annotations and assessments are publicly and freely available for use, and all data is non-personal and anonymized.

\HARD\ develops automatic and manual criteria for categorising complex queries, which is applicable when constructing future datasets. We use \HARD\ to compare the overall system effectiveness of systems in the TREC 2020 DL track. We find significant differences in system ordering and an overall reduction in effectiveness (headroom for future research). This resource represents an important step towards more challenging datasets for passage and document ranking.

\section{Future Work}
\label{sec:future-work}
The authors plan future work to remove duplicates found in the MS MARCO collection, add explicit long descriptions to topics to remove ambiguity, and add an entity ranking task to complement the current document/passage ranking tasks. 

\section{Acknowledgements}
\label{sec:ack}

This work is supported by the Engineering and Physical Sciences Research Council grant EP/V025708/1, the 2019 Bloomberg Data Science Research Grant, and the TensorFlow Research Cloud.

\bibliographystyle{ACM-Reference-Format}
\bibliography{foo}

\end{document}